\documentclass[superscriptaddress, twocolumn,showpacs,preprintnumbers,amsmath,amssymb]{revtex4}

\usepackage{graphicx}
\usepackage{dcolumn}
\usepackage{bm}
\usepackage{color}

\begin{document}

\preprint{APS/123-QED}

\title{Long-lived electron spin coherence in CdSe/ZnSSe self-assembled quantum dots}

 \author{M. Syperek}
 \affiliation{Experimentelle Physik 2, Technische Universit\"at Dortmund, 44221 Dortmund, Germany}
 \affiliation{Institute of Physics, Wroc\l{}aw University of Technology, 50-370 Wroc\l{}aw, Poland}
 \author{D. R. Yakovlev}
 \affiliation{Experimentelle Physik 2, Technische Universit\"at Dortmund, 44221 Dortmund, Germany}
 \affiliation{Ioffe Physical-Technical Institute, Russian Academy of Sciences, 194021 St.Petersburg,  Russia}
 \author{I. A. Yugova}
 \affiliation{Experimentelle Physik 2, Technische Universit\"at Dortmund, 44221 Dortmund, Germany}
 \affiliation{Physical Faculty of St.Petersburg State University,
198504 St.~Petersburg, Russia}
 \author{J. Misiewicz}
 \affiliation{Institute of Physics, Wroc\l{}aw University of Technology, 50-370 Wroc\l{}aw, Poland}
 \author{I. V. Sedova}
\affiliation{Ioffe Physical-Technical Institute, Russian Academy of Sciences, 194021 St.Petersburg,  Russia}
 \author{S. V. Sorokin}
\affiliation{Ioffe Physical-Technical Institute, Russian Academy of Sciences, 194021 St.Petersburg,  Russia}
 \author{A. A. Toropov}
 \affiliation{Ioffe Physical-Technical Institute, Russian Academy of Sciences, 194021 St.Petersburg,  Russia}
 \author{S. V. Ivanov}
\affiliation{Ioffe Physical-Technical Institute, Russian Academy of Sciences, 194021 St.Petersburg,  Russia}
 \author{M. Bayer}
 \affiliation{Experimentelle Physik 2, Technische Universit\"at Dortmund, 44221 Dortmund, Germany}

\date{\today}

\begin{abstract}
The electron spin coherence in \textit{n}-doped and undoped,
self-assembled CdSe/Zn(S,Se) quantum dots has been studied by
time-resolved pump-probe Kerr rotation. Long-lived spin coherence
persisting up to 13 ns after spin orientation has been found in the
\textit{n}-doped quantum dots, outlasting significantly the
lifetimes of charge neutral and negatively charged excitons of
$350-530$~ps.  The electron spin dephasing time as long as 5.6 ns has been measured in a magnetic field of 0.25 T. Hyperfine interaction of resident electrons with  a nuclear spin fluctuations is suggested as the main limiting factor for the dephasing time. The efficiency of this mechanism in II-VI and III-V quantum dots is analyzed.
\end{abstract}


\maketitle
\section{Introduction}

The electron spin coherence in semiconductor quantum dots (QDs) has
been extensively studied experimentally and theoretically  in recent
years.\cite{Spinbook, QBbook, Gabibook} This activity is driven by
prospective applications of the electron spin in quantum information
processing requiring methods of initialization, manipulation,
storage, and readout of spin coherence. A vital problem in this
respect concerns the coupling of the electron spin to its
environment, deteriorating the spin dynamics. The strength of this
interaction determines the transverse spin relaxation time
$\emph{T}_2$, limiting coherent state manipulation. In QDs the
electron wave function is strongly spatially localized by the
three-dimensional confinement potential. On one hand this leads to
discrete energy levels and makes the electron spin weakly
susceptible to the main spin relaxation mechanisms in bulk
semiconductors related to the spin-orbit interaction.\cite{Meyer84}
On the other hand it enhances the Fermi contact hyperfine
interaction, which therefore is considered as main relaxation
mechanism for QD electron spins. \cite{Merkulov02, Khaetskii03} The
hyperfine interaction is controlled by: (1) the nuclear isotope
composition, (2) the hyperfine coupling constants, (ii) the number
of lattice nuclei, and (iv) the nuclear spin magnitudes. Therefore,
a proper choice of the QD semiconductor material may minimize this
interaction, potentially prolonging the transverse spin relaxation
time. In this respect II-VI semiconductors are an interesting
alternative to III-V materials. For example, in CdSe the abundance
of nuclear isotopes with nonzero spin is 7.6\% for Se and 25\% for
Cd only, compared with the 100\% for In, Ga and As in the widely
studied (In,Ga)As dots.\cite{Ch4} Therefore, the electron spin
dephasing by the hyperfine interaction should be considerably weaker
in CdSe dots.

Coherent electron spin dynamics was studied by time-resolved Faraday
rotation for colloidal CdSe QDs~\cite{Gupta99, Gupta02, Ouyang03,
Stern05, Li06}  and Cd(S,Se) QDs in semiconductor-doped
glasses\cite{Gupta01}, both of wurtzite-type. Longitudinal spin
relaxation times of $\emph{T}_1 \sim 20$~$\mu$s and transverse spin
dephasing times $T_2^* \sim 3$~ns were reported.\cite{Gupta99,
Gupta02} The measured Faraday rotation traces show complicated
quantum beat patterns due to the QD-anisotropy typical for wurtzite
structures (leading to a complex exciton fine structure) and also
due to the random dot orientation in the ensemble, so that the
optical selection rules become undefined. These factors complicate
the theoretical description of coherent spin dynamics in colloidal
QDs, in addition to the influence of surface states.\cite{Gupta02}

Some of these complications become resolved in epitaxially-grown
self-assembled QDs. First, the influence of surface states is
eliminated. Second, modulation doping allows one to fabricate
charged QDs with a single resident electron per dot.\cite{Ch6} The
spin dynamics of resident electrons is not limited by radiative
exciton recombination as in empty QDs. Resident electron spin
coherence can be generated via an intermediate trion state, as
suggested theoretically\cite{Shabaev03} and demonstrated
experimentally in (In,Ga)As/GaAs QDs.\cite{Greilich06, Ch6} Another
advantage of self-assembled QDs is their well-defined
crystallographic orientation on the substrate, from which optical
selection rules are established. Also the crystal structure can be
controlled by the choice of substrate material, e.g., grown on a
GaAs substrate, CdSe QDs have cubic symmetry instead of wurtzite
symmetry, simplifying the exciton fine structure.

In this paper we report on time-resolved pump-probe Kerr rotation
(KR) studies of the electron spin coherence in \textit{n}-doped and
undoped self-assembled CdSe/Zn(S,Se) QDs with a cubic crystal
structure. We observe long-lived spin beats related to the coherent
precession of electron spins about a transverse magnetic field. The
coherence persists over times exceeding the interval between
subsequent laser pulses of $\sim$13~ns. Analysis of the KR traces as
function of magnetic field gives information about dephasing times
and $g$ factors of the resident QD electrons. Also the temperature
dependence of the dephasing time was investigated, and the results
are compared to literature data, especially for colloidal wurtzite
CdSe QDs.

\section{Experiment}

The investigated CdSe/Zn(S,Se) QD multilayer structures were
pseudomorphically grown by molecular-beam epitaxy on a GaAs:Si
buffer layer on top of an \textit{n}-type GaAs (001) substrate.
First, a 1.5~$\mu$m-thick \textit{n}-type ZnMgSSe:Cl wide-gap
confining layer ($E_g=2.96$~eV at $T=77$~K) was deposited to
separate the QD region from the GaAs buffer.  Then, ten periods of
CdSe QD layers (each having a nominal thickness of 2.5 ML with a
self-assembled dot density of $5-8 \times 10^{10}$~cm$^{-2}$) were
grown, each layer sandwiched between Zn(S,Se)/ZnSe short-period
superlattice (SL) barriers with a thickness of 50~nm. The
compressive strain induced by the CdSe QDs is compensated by the
tensile strain intentionally generated in the SL barriers, which
provides similar growth conditions for each QD layer. Finally, the
entire structure was capped by a 30~nm-thick ZnMgSSe layer. Two
structures were fabricated: a reference sample with nominally
undoped QDs and a sample with \textit{n}-doped QDs. In the
\textit{n}-doped structure, the middle part of each SL barrier with
a thickness of 17 nm was doped by Cl with a concentration resulting
in an electron sheet density of $(3-5) \times 10^{11}$~cm$^{-2}$. We
expect that only $10-20$~\% of the electrons are transferred to the
QDs through the undoped 17~nm-thick SL spacers. From these numbers
we expect that the QD ensemble represents mostly a mix of
charge-neutral and singly-charged QDs, but there may be also a small
fraction of doubly or higher charged dots.

The pulsed laser source used in the experiments was an optical
parametric oscillator pumped by a mode-locked Ti:Sapphire laser. The
laser system generates pulses with a duration of 1.6~ps at a
repetition rate of 75.6~MHz, corresponding to 13.2~ns pulse
separation. The photon energy could be tuned in the range
$2.38-2.47$~eV. The samples were held in a helium bath cryostat
equipped with a superconducting split-coil magnet. Magnetic fields
$\textbf{B}$ up to 7~T were applied normal to the structure growth
\emph{z}-axis (Voigt geometry, $\textbf{B} \bot \textbf{z}$). The
sample temperature was varied from 1.6 up to 50~K.

In the KR experiment the laser beam was split into a circularly
polarized pump and a linearly polarized probe. Both beams were
focused on the sample with a spot diameter of $\sim$150~$\mu$m. To
avoid dynamic nuclear polarization, the helicity of the pump beam
was modulated at 50~kHz rate using a photoelastic modulator.  The
pump pulse excites carriers with spins polarized along the
\emph{z}-axis. The subsequent coherent spin dynamics in form of
Larmor precession about $\textbf{B}$ is measured by the rotation
angle $\Theta_{KR}$ of the polarization plane of the probe. To
detect $\Theta_{KR}$, a homodyne technique based on phase-sensitive
balanced detection was used.\cite{Ch6}

The spectrally-resolved recombination dynamics of the optically
excited electron-hole pairs could be detected in time-resolved
photoluminescence (TRPL). For that purpose the samples were mounted
in a helium flow cryostat at $T=8$~K. The PL emission was dispersed
by a 0.5~m spectrometer and detected by a synchroscan streak camera
with an S20 photocathode. The overall time resolution of this setup
was 15~ps.

\section{Results and discussion}

\begin{figure}[b]
\includegraphics[width=7 cm]{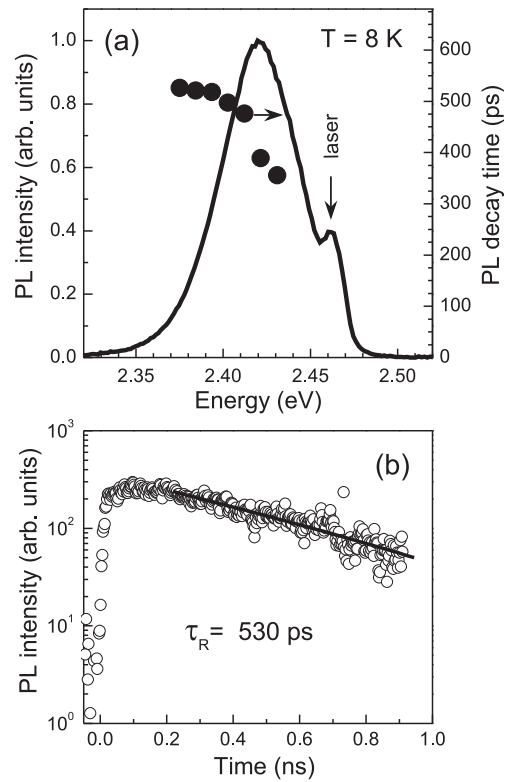}
\caption{\label{fig:epsart}(a) Photoluminescence spectrum (line) and
photoluminescence decay time dispersion (circles) of the
\textit{n}-doped CdSe/Zn(S,Se) QDs measured for quasi-resonant
excitation at 2.46~eV photon energy. (b) Example of
the PL intensity decay at 2.38~eV after pulsed excitation, plotted
on a logarithmic scale: circles - experiment, line - fit.}
\label{fig:1}
\end{figure}

The low temperature photoluminescence spectrum of the
\textit{n}-doped CdSe/Zn(S,Se) QDs under quasi-resonant excitation
(see arrow indicating the laser photon energy) is shown in
Fig.~\ref{fig:1}(a). The QD emission due to radiative recombination
of negatively charged excitons (trions) in singly-charged dots and
of charge neutral excitons in empty QDs is centered at 2.42~eV. The
emission is characterized by considerable spectral broadening of
$\sim$50~meV due to QD inhomogeneities. A similar PL spectrum is
obtained for the undoped QDs (not shown).

The exciton recombination dynamics was measured by TRPL. An example
of the PL decay at 2.38~eV is shown in Fig.~\ref{fig:1}(b). The
decay can be fitted well by a single exponential as shown by the
solid line with a characteristic time of $\tau_{R}=530$~ps. The PL
decay time varies across the emission band: $\tau_{R}$ increases
from 350~ps on the high energy flank at 2.43~eV to 530~ps on the low
energy side at 2.38~eV, see circles in Fig.~\ref{fig:1}(a). The
observed systematic increase of the decay  times with decreasing
transition  energy  is  in agreement with previous measurements on
epitaxially-grown  CdSe/ZnSe  QDs,  both  for ensembles\cite{Gindel}
and single dots\cite{Patton}.  The  increase in  the  exciton
radiative  lifetime with increasing localization energy can be
explained by a decrease of the exciton coherence volume in real
space and, hence, its increased spread in K space\cite{Sugawara}. We
note here, that in the studied structures the effective barrier
energy provided by the superlattice minibands is 2.84~eV, which
implies the strong confinement regime for the confined electron and
hole wavefunctions.

\begin{figure}[t]
\includegraphics[width=7.5 cm]{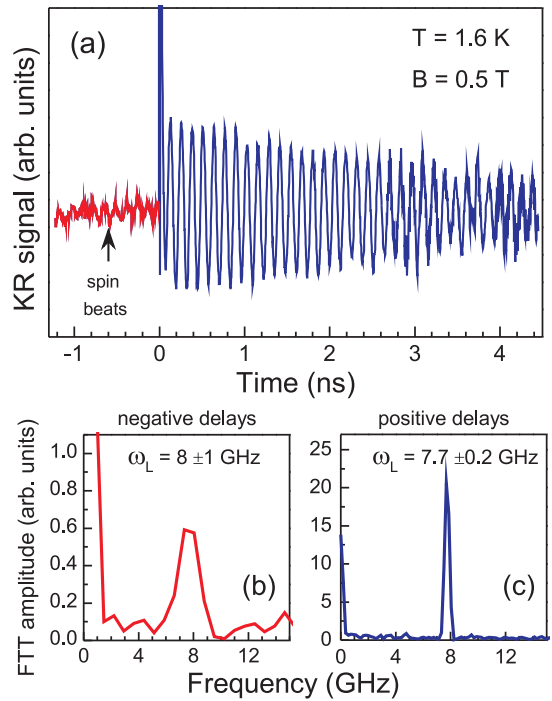}
\caption{\label{fig:epsart} (color online) (a) Time-resolved Kerr
rotation trace of the \textit{n}-doped CdSe/Zn(S,Se) QDs \textit{vs}
time delay between pump and probe pulses. FFT amplitudes of KR
signals analyzed for negative (b) and positive (c) time delays. Pump
power 60~W/cm$^{2}$, probe power 5~W/cm$^{2}$, laser photon energy
2.44~eV.} \label{fig:4}
\end{figure}

The spin dynamics was investigated by pump-probe Kerr rotation.
Figure~\ref{fig:4}(a) shows the KR trace for the \textit{n}-doped
QDs measured at $B=0.5$~T. The laser photon energy was tuned to
2.44~eV for resonant excitation of the trion and exciton states in
the QDs. The KR signal shows features typical for structures
containing singly-charged QDs or quantum wells with a low density
electron gas.\cite{Ch6} At positive time delays relative to the pump
pulse at time zero, a periodically oscillating signal with
exponentially decreasing amplitude for increasing delays is
observed. Remarkably the signal amplitude is still strong at delays
of 4.5~ns, i.e., at times strongly exceeding the characteristic
trion lifetimes, that do not exceed 0.6~ns, see Fig.~\ref{fig:1}(a).
This allows us to conclude that the KR signal is dominated by the
contribution of the resident electrons in singly-charged QDs, whose
lifetime is not limited by recombination. A closer look at negative
time delays, see the red line in Fig.~\ref{fig:4}(a), reveals spin
beats with rather small amplitude.

The spin precession frequencies can be evaluated by Fast Fourier
Transformation (FFT) as shown in Figs.~\ref{fig:4}(b) and
\ref{fig:4}(c). The precession frequencies for negative and positive
delays are close to each other: $\omega_L=$8$\pm$1~GHz and 7.7$\pm$0.2~GHz,
respectively. This small difference in $\omega_L$ is within the
experimental error due to the noisy low amplitude signal at negative
delays, which results in a significant broadening of the FFT
spectrum and a strong noise peak around zero frequency. These
results also support the incomplete dephasing of the electron spin
coherence within a laser pulse interval of 13.2~ns.

\begin{figure}[t]
\includegraphics[width=8 cm]{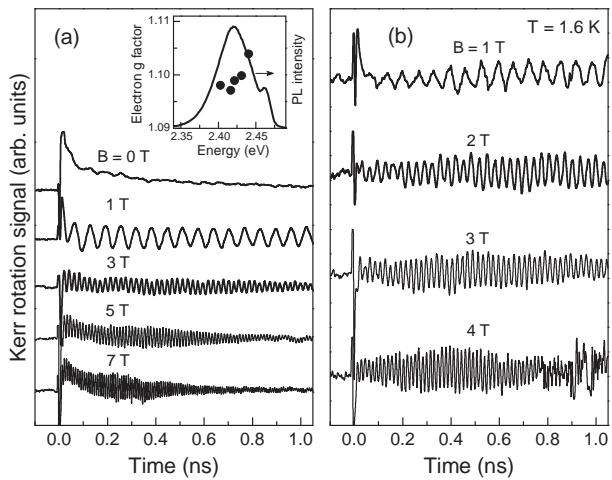}
\caption{\label{fig:epsart} KR signals in \textit{n}-doped (a), and
undoped (b) CdSe/Zn(S,Se) QDs at different magnetic fields. Pump
power 60~W/cm$^{2}$, probe power 5~W/cm$^{2}$, laser photon energy
2.44~eV. The inset shows the electron $g$ factor dispersion across
the PL band of the \textit{n}-doped QD ensemble (circles) together
with the PL spectrum (line) excited at 2.46~eV.} \label{fig:2}
\end{figure}

We consider now the Kerr rotation signals measured on the
\textit{n}-doped CdSe/Zn(S,Se) QDs for different magnetic fields
in the delay range up to 1~ns, see Fig.~\ref{fig:2}(a). At zero magnetic field the light
induced spin polarization decays with three components. Two of them
have characteristic times of 30 and 300~ps. The third, longest
lasting component exceeds the lifetimes of excitons and trions and
can be assigned to the spin relaxation of the resident electrons,
while the shorter lived components correspond to relaxation
processes within the trion complexes, which are limited by either
spin relaxation or recombination.  The signals for $B>0$~T are
dominated by long-lived oscillations with a single frequency
corresponding to the resident electron Larmor frequency. Also a
second non-precessing component is present at relatively short delay
times up to 250~ps. This time range coincides well with the trion and exciton lifetimes,
so that the non-precessing component can be assigned to the spin
dynamics of these exciton complexes.\cite{Yug07}

The KR signals from nominally undoped QDs are shown in
Fig.~\ref{fig:2}(b). Their amplitudes are about 20 times weaker, but
otherwise their appearance is qualitatively similar to that of the
\textit{n}-doped sample. The signals are dominated by long-lived
electron spin beats, which can be explained by unintentional doping
of a fraction of dots in the ensemble. No strong hole contribution
is seen in this sample at short delays.

In the following we focus on the KR signals of the \textit{n}-doped
QD sample and analyze them for delays exceeding $200-300$~ps to
study the spin coherence of resident electrons only. For this
analysis we used an exponentially damped harmonic~\cite{Ch6} to fit
the KR signals and thereby obtain the spin beat frequency and the
spin dephasing time:
\begin{equation}
\label{KR_exp}
\Theta_{KR}(t) = {\Theta_{KR}(0)\exp \left( - \frac{t}{{T}^*_{2}}\ \right)\cos\left({\omega_L t}\right)}.
\end{equation}
Here $\Theta_{KR}(0)$ is the KR angle at the moment of pump pulse
arrival $t=0$~ps, $T_2^*$ is the spin dephasing time, and
$\omega_L$ is the Larmor spin precession frequency~\cite{comment1}.

First, we discuss the experimental data on the in-plain
Land$\acute{e}$ factor $g_e$ of the electron. From the $\omega_L$
obtained through the fitting, the $g_{e}$ can be derived by:
$g_e=\hbar\omega_L/(\mu_B B)$. The inset in Fig.~\ref{fig:2}(a)
shows the dependence of $g_e$ on the QD emission energy. $g_e$ has a
value of about 1.1 with a weak dispersion across the PL band originating from the $g$ factor dependence on the band gap width\cite{Yugova07b}. Our
experimental technique is not sensitive to the sign of $g_e$, but it
is known that $g_e$ is positive in CdSe, see Ref.~\cite{Karimov00}
and references therein.

\begin{figure}[t]
\includegraphics[width=7.5 cm]{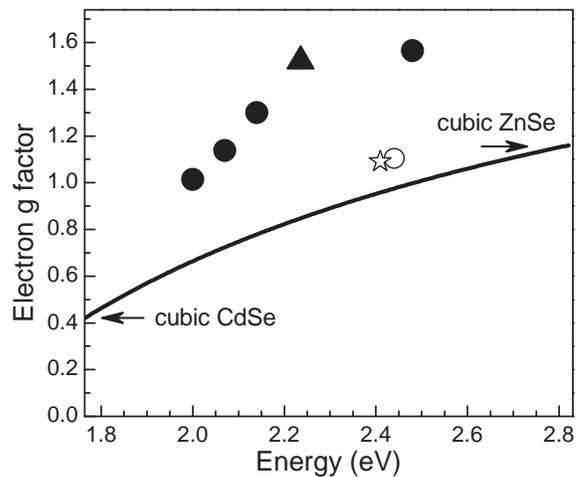}
\caption{\label{fig:epsart} Electron $g$ factor as function of
optical transition energy: \textit{n}-doped, self-assembled
CdSe/Zn(S,Se) QDs (open circle), self-assembled CdSe/ZnSe QDs (open
star) \cite{Puls99}, wurtzite colloidal CdSe QDs (closed circles)
\cite{Gupta02}, and wurtzite Cd(S,Se) nanocrystals in glass (closed
triangle) \cite{Gupta01}. Values for cubic bulk materials are shown
by arrows: $g_e=0.42$ for CdSe and 1.15 for ZnSe.\cite{Karimov00}
The dispersion for cubic (Zn,Cd)Se calculated after
Ref.~\cite{Karimov00} is shown by the solid line.} \label{fig:5}
\end{figure}

In Fig.~\ref{fig:5} we compare the $g_e$ value  for the studied QDs
with those reported in literature. We chose a presentation form in
which the $g_e$ is given in dependence on the energy of optical
transition (corresponding to the energy gap), as this is the most
straightforward way to compare the experimental results with
theoretical predictions based on the Roth approach for
bulk~\cite{Roth59}. It was shown that this approach can be
successfully extended also to quantum well and quantum dot
heterostructures, where the quantum confinement energies have to be
added to the bulk band gap energy to obtain the optical transition
energies, see e.g. Refs.~\cite{Sirenko97, Yugova07b, Gupta02}. In
Fig.~\ref{fig:5} the experimental data for cubic self-organized QDs
are shown by the open symbols, and for wurtzite QDs they are given
by the closed symbols. Our result for CdSe/Zn(S,Se) dots (open
circle) is in good agreement with the one for cubic CdSe/ZnSe dots
(open star).\cite{Puls99} They are also in reasonable agreement with
the calculations in Ref.~\cite{Karimov00} for cubic bulk (Zn,Cd)Se,
shown by the solid line. These calculations, in turn, agree well
with experimental data for (Zn,Cd)Se epilayers grown on GaAs
substrates.\cite{Karimov00} However, the results for the wurtzite
CdSe dots shown by the closed symbols deviate considerably from that
data for cubic dots. This difference is due to differences in the
band parameters for wurtzite and cubic materials, which can be also
well accounted for in the Roth approach.\cite{Gupta02}

\begin{figure}[b]
\includegraphics[width=7.5 cm]{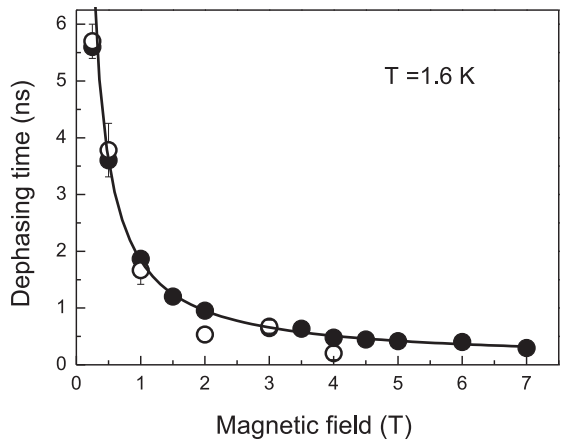}
\caption{\label{fig:epsart} Spin dephasing time $T^*_2$ \textit{vs}
transverse magnetic field for the \textit{n}-doped (closed circles)
and the undoped (open circles) self-assembled CdSe/Zn(S,Se) QDs.} \label{fig:3}
\end{figure}

The spin dephasing times for the resident electrons evaluated from
the experimental data in Fig.~\ref{fig:2} are presented in
Fig.~\ref{fig:3} as function of magnetic field strength. The results
for \textit{n}-doped and undoped QDs coincide well and show a strong
decrease of the $T^*_2$ time from 5.6~ns at $B=0.25$~T down to
300~ps at $B=7$~T.  The dephasing time is contributed by the
coherence time of individual spins, $T_2$, and the inhomogeneous
dephasing time caused by the inhomogeneity of the spin ensemble
$T_2^{inh}$ [\onlinecite{Ch6}]:
\begin{equation}
\frac{1}{T^*_2} = \frac{1}{T_2} + \frac{1}{T_2^{inh}}.
\end{equation}
The $T_2^{inh}$ time is contributed, e.g., by the nuclear spin
fluctuations (dominant at very low fields) and by the spread of the
electron Larmor frequencies (dominant at high fields). The latter
provides the magnetic field dependent contribution to the
$T_2^{inh}$ time, and therefore to the $T^*_2$ time:
\begin{equation}
\frac{1}{T_2^{inh}(B)}= \frac{1}{T_2^{inh}(0)} + \frac{\Delta g_e \mu_B B}{\hbar}.
\end{equation}
Here $\Delta g_e$ is the spread of the electron $g$ factor for the
optically excited spin ensemble. One can see from Fig.~\ref{fig:3}
that the experimental data can be well described by a $1/B$
dependence for $B > 0.25$ T, as shown by the solid line assuming
a $\Delta g_e=0.0055$. This demonstrates that the dephasing time for
not too small magnetic fields is governed by the spread of the
electron spin Larmor precession frequency. A similar behavior has
been reported for colloidal CdSe QDs.\cite{Gupta99, Gupta02} Note
that this spread $\Delta g_e=0.0055$ is less that 0.6~\% of the mean
$g_e$ value.

The dephasing time $T^*_2=5.6$~ns at 0.25~T is one of the longest
reported so far for QDs. It is larger than the dephasing
times reported for colloidal CdSe QDs of 3~ns at $B=0.25$~T
\cite{Gupta99}, and than $T^*_2 \sim 2$~ns at $B=0.2$~T in
(In,Ga)As/GaAs QDs.\cite{Greilich_Science313}

Around zero applied magnetic field the spin dephasing in quantum
dots is expected to be determined by the electron interaction with
the nuclear spin fluctuations at low temperatures. According to
Ref.~\cite{Merkulov02} the electron spin dephasing time due to this
interaction can be calculated according (see Appendix for details):
\begin{equation}
\label{t_nucl}
T_2^*=\sqrt{2} T_{\Delta}=\hbar \sqrt{\frac{3N_L}{2 \sum_j I_j(I_j+1)\tilde{A_j^2}y_j}} .
\end{equation}
Here the sum runs over all types of nuclear isotopes in the dots,
$N_L$ is the number of nuclei in the QD volume, and $I_j$ is the
nuclear spin. $\tilde{A_j}$ is the hyperfine constant, which is
taken for a unit cell with two nuclei, and $y_j$ is the probability
to find the particular type of nuclear isotope (see Appendix for
details). Using this equation we can evaluate the electron spin
dephasing time in CdSe dots. To estimate the number of nuclei we
used the geometrical volume of the dot, assumed to be of cylindrical
shape (QD diameter of $4-6$~nm and height of $1.4-2.1$~nm), and used
hyperfine coupling constants of -37.4~$\mu$eV, -39.1~$\mu$eV, and
33.6~$\mu$eV for the stable isotopes of $^{111}$Cd, $^{113}$Cd, and
$^{77}$Se, respectively, each having spin 1/2 (see Table~I and
Appendix for details). This gives a $N_L$ in the range from 630 to
2090 nuclei and leads to a dephasing time of $1-2$~ns depending on
the QD size. For better agreement with the measured 5.6~ns we have
to assume that either the QD volume is an order of magnitude larger,
which can be achieved by assuming, for example, a height of 2.1~nm
and a diameter of 17~nm or a height of 4~nm and a diameter of 12~nm
(see Appendix), or the interaction with the nuclei is weaker, e.g., due to deviation of the real situation from the modeled one in the exchange-box approximation.

If the QDs would have the same size, the nuclei-induced spin
dephasing would be much more efficient in GaAs and (In,Ga)As dots
compared to CdSe dots, because the nuclear spins in III-V dots are
considerably larger: 9/2 for In and 3/2 for As and Ga compared to
1/2 for Cd and Se. A comparison of the relevant parameters for CdSe
and (In,Ga)As dots is given in Table~I. In addition, $T_2^{\star}$
scales inversely with the number of nuclei with non-zero spin. The
abundance of such isotopes in (In,Ga)As is 100\% for all nuclear
species, while it is only 25\% for Cd and 7.6\% for Se. This
shortens the dephasing time in III-V QDs further. As a result, for
dots of the same size one would expect the electron spin dephasing
times to be about 10 times shorter in (In,Ga)As than in CdSe dots,
see Fig.~7(a). In practice, (In,Ga)As/GaAs QDs grown by
molecular-beam epitaxy have, however, considerably larger sizes than
CdSe self-organized dots. This reduces the difference in the number
of non-zero spin nuclei, and therefore also the difference in
dephasing times.

It is important to mention though that numerical estimates based on
Eq.~(\ref{t_nucl}) give smaller values of $T_2^*$ than
experimentally observed for both CdSe and (In,Ga)As QDs. Most
probably this is related to the limitations of the used approach,
which is based on the "box model" assuming a constant density of the
electron envelope wave function at all nuclear sites within the QD.
Also possible effects of nuclear polarization, e.g. nuclear
frequency focusing \cite{focusing}, are beyond the model
consideration. Accounting for these effects would require detailed
information on the QD shape and substantial computational efforts,
both of which go beyond the scope of the present paper.

\begin{figure}[t]
\includegraphics[width=7.5 cm]{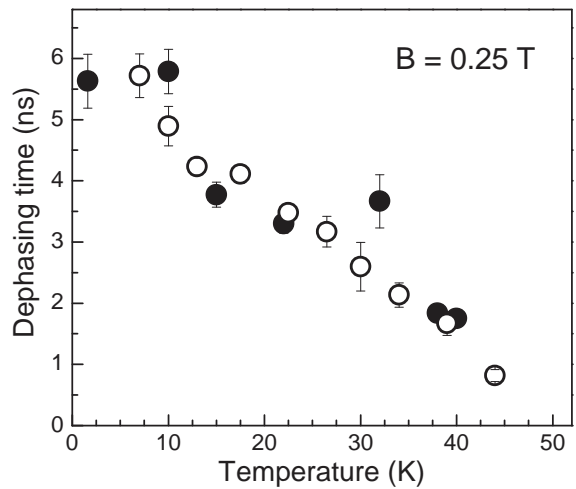}
\caption{\label{fig:epsart} Temperature dependence of the dephasing
time $T_2^*$ for the \textit{n}-doped (closed circles) and the
undoped (open circles) QDs. Pump power 60~W/cm$^{2}$, probe power
5~W/cm$^{2}$, photon energy 2.43~eV. } \label{fig:6}
\end{figure}

We examined also the temperature effect on the KR signal.
Figure~\ref{fig:6} shows how the dephasing time is changing with
increasing temperature in a weak magnetic field of $B=0.25$~T. The
$T^*_2$ time of the resident electrons decreases monotonically from
5.6~ns at $T=1.6$~K ($k_BT=0.14$~meV) down to 0.8~ns at $T=44$~K
($k_BT=3.8$~meV).

While a strong temperature dependence of transverse electron
spin relaxation times has been reported already for self-assembled
QDs, e.g., (In,Ga)As/GaAs \cite{Hernandez08} and InP/(In,Ga)P QDs
\cite{Masumoto06}, the
underlying mechanisms are not trivial and are still not fully
understood. They are not related to thermal escape of resident
electrons, as the dot confinement potential considerably exceeds the
characteristic thermal energies. Also $\Delta g_e$ is not expected
to vary strongly in the scanned temperature range, so that
shortening due to related inhomogeneities can be most likely
excluded.

Consequently, the shortening of the dephasing time $T^*_2$ with
increasing temperature can be related to the decrease of $T_2$.
Spin-orbit related dephasing is expected to be too weak to account
for this shortening (see Ref.~\cite{Hernandez08} and references
therein). An effective spin relaxation mechanism providing electron
spin decoherence in a quantum dot has been proposed in
Ref.~\cite{Semenov04}. It suggests phonon-mediated fluctuations in
the electron spin precession, which are caused by the modulation of
the longitudinal electron $g$ factor and the hyperfine field.
Indications to that end were observed also for self-assembled
(In,Ga)As/GaAs QDs~\cite{Hernandez08}, but more detailed studies are
required to clarify the role of this mechanism for CdSe/Zn(S,Se)
QDs.

\section{Conclusions}

The coherent spin dynamics of resident electrons in self-assembled
CdSe/Zn(S,Se) QDs have been investigated by time-resolved
photoluminescence and Kerr rotation. Long-lived electron spin
coherence detectable up to 13~ns time delays has been found. The
spin dephasing time of 5.6~ns at $B=0.25$~T is one of the longest
reported so far for CdSe and (In,Ga)As QDs. The magnetic field
dependence of the dephasing time follows well a $1/B$ dependence
reflecting a relatively small spread of the electron $g$ factor
$\Delta g_e=0.0055$ around the mean value of $g_e=1.10$ within the
QD ensemble excited by the laser pulse with a spectral width of
about a meV.

\section{Acknowledgments}

The work was supported by the Deutsche Forschungsgemeinschaft, the
EU Seventh Framework Programme (Grant No. 237252, Spin-optronics),
the Russian Foundation for Basic Research (Grant No.
09-02-12406-ofi-m) and the Polish MNiSW (Grant No. N202179538). M.S. is thankful to the Foundation of Polish
Science within the "START" programme. The research stay of I.A.Y. in
Dortmund is supported by the Alexander von Humboldt Foundation,
Bonn.

\appendix
\section{}

Here we give details of the evaluation of the nuclear spin
fluctuation contribution to the dephasing time for an electron spin
ensemble in singly charged QDs.  Note, that for the hyperfine
constants $A_j$ of a particular nuclear species different values can
be found in literature
\cite{Paget77,Merkulov02,Nakamura79,Testelin09}. The reason for that
are differences in defining the crystal unit cell for which the
$A_j$ are calculated. Further the isotope abundance was either
included \cite{Merkulov02} or not \cite{Testelin09} in $A_j$.

The Hamiltonian of hypefine Fermi contact interaction between
electron and nuclear spins is:\cite{Abragam}
\begin{equation}
\hat{H}_{hf}=\sum_k a_k(\hat{{\bm S}} \hat{{\bm I}_k}),
\end{equation}
where the sum goes over all nuclei, $\hat{{\bm S}}$ is the spin
operator of the electron, $\hat{{\bm I}_k}$ is the spin operator of
the $k$-th nucleus, $a_k=v_0A_k|\psi({\bm R}_k)|^2$, $v_0$ is the
unit cell volume, $A_k$ is the hyperfine constant:
\begin{equation}
A_k=\frac{16\pi \mu_B \mu_k | u_c({\bm R}_k)|^2 }{3I_k},
\end{equation}
with $\mu_k$ and $I_k$ being the magnetic moment and spin of the
$k$-th nucleus, the Bohr magneton $\mu_B$, the electron envelope
wave function $\psi({\bm R}_k)$ at the $k$-th nucleus, and $u_c({\bm
R}_k)$ being the electron Bloch function at the $k$-th nucleus. We
write the electron wave function in the form  $\Psi ({\bm R}_k) =
v_0 \psi({\bm R}_k) u_c({\bm R}_k)$, as in
Ref.~[\onlinecite{Merkulov02}]. The normalization conditions are
$\int_V|\psi({\bm R}_k)|^2 dv =1$ and $\int_{v_0}|u_c({\bm R}_k)|^2
dv=1$. With this definition $|u_c({\bm R}_k)|^2 \sim 1/v_0$, and
therefore one can see from Eq.~(A2) that the $A_k$ are also $\sim
1/v_0$.

To estimate the electron spin dephasing time we need to calculate
the dispersion of the hyperfine field distribution $\Delta_B^2$. We
assume that the fluctuating nuclear field is sum of the independent
contributions of the different types of nuclei in the dots.
Therefore, the dispersion $\Delta_B^2$ has to be calculated as sum
of the dispersions from different contributions:
\begin{equation}
\Delta_B^2=\sum_{j} (\Delta_B^{j})^2,
\end{equation}
where the sum runs over the nuclei types. Each individual
contribution is:
\begin{equation}
(\Delta_B^{j})^2= \frac{2v_0^2
I_{j}(I_{j}+1)}{3(g_e\mu_B)^2}\sum_{j, \xi} (A_{j, \xi})^2|\psi({\bm
R}_{j, \xi})|^4,
\end{equation}
with the sum with $\xi$ running over all nuclei of the same type $j$
in a QD. We introduce the dimensionless electron density 
$\eta_j=| u_c({\bm R}_j)|^2v_0$ which has its maximum at the $j$-th nucleus, and also the
number of nuclei in the volume of electron localization $V_L$:
\begin{equation}
N_L=\frac{nV_L}{v_0},
\end{equation}
where $V_L=\left(\int dv|\psi({\bm r})|^4 \right)^{-1}$ and $n$ is
the number of nuclei in the unit cell. For the unit cell with two
nuclei ($n=2$): $N_L=2V_L / \tilde{v}_0$,
$\tilde{A_j} \equiv 16\pi \mu_B \mu_j \eta_{j}/(3I_j\tilde{v}_0)$, 
where $\tilde{v}_0$ is the unit cell
volume. For an unstrained cell $\tilde{v}_0=a_0^3/4$, where $a_0$ is
the lattice constant.

One can replace the sum over unit cells by an integration and
evaluate the sum for all nuclei of the same type in a cell:
\begin{equation}
(\Delta_B^{j})^2= \frac{4}{3(g_e\mu_B)^2}\frac{1}{N_L} I_{j}(I_{j}+1)\tilde{A_j^2}y_{j}
\end{equation}
where $y_{j}$ is the probability to find a nucleus of type $j$.
$y_j=x_{j}\varkappa_j$, where $\varkappa_j$ is the isotope abundance
($\varkappa \in [0,1]$ as in Ref.~[\onlinecite{Paget77}]) and $x_j$
is the concentration of the substitution ions in a ternary alloy,
e.g., in In$_x$Ga$_{1-x}$As. For binary materials $x_j=1$.

Thereby we obtain an electron spin dephasing time of:
\begin{equation}
T_2^*=\hbar \sqrt{\frac{3N_L}{2 \sum_j I_j(I_j+1)\tilde{A_j^2}y_j}} .
\end{equation}
Here the sum is over the nuclear species. The experimental value of
spin dephasing time corresponds approximately to the half width at
half maximum of the normal distribution \cite{comment1}. Therefore
the relation to the measured $T_2^*$ can be made through
$T_2^*=\sqrt{2}T_{\Delta}$, where $T_{\Delta}$ is equivalent to the
time introduced by Eq.(11) in Ref.~\cite{Merkulov02}. The difference
in coefficients is accounted for by an other choice of the unit cell
with 8 atoms in Ref.~\cite{Merkulov02}, instead of 2 atoms used in
our paper.

The values of $\eta_j$ for nuclei Ga, As, In, Cd, and Se are given
in Table~I. The data for Ga, As, In, and Cd were taken from
Refs.~\cite{Paget77,Nakamura79,Gueron}.  In these papers the unit
cell containing two nuclei has been chosen, i.e., $v_0=a_0^3/4$. To
get $\eta_j$ we used $a_0$(GaAs)$=0.565$~nm \cite{Landolt},
$a_0$(InSb)$=0.648$~nm \cite{Gueron}, and $a_0$(CdTe)$=0.648$~nm
\cite{Landolt26}. For Se we did not find a value of
$\eta_\mathrm{Se}$ in literature. 
We assume that the bonds in CdSe
are covalent and take $\eta_\mathrm{Se}=\eta_\mathrm{Cd}$ for our
estimation, as was done in Ref.~\cite{Nakamura79}. Values of
$\tilde{A_j}$ calculated with unit cells with two atoms (with
$a_0$(GaAs)$ = 0.565$~nm for In, Ga, As nuclei and with
$a_0$(CdSe)$=0.6077$~nm for Cd, Se nuclei) are given in Table~I.

Numerical calculations give $T_2^* \approx 43.7 \sqrt{N_L}$~[ps] for
CdSe (with $a_0$(CdSe)$=0.6077$~nm \cite{Landolt26}). The equivalent
dependence for GaAs is $T_2^* \approx 6.63 \sqrt{N_L}$~[ps] and for
In$_x$Ga$_{1-x}$As: $T_2^* \approx 6.63
\sqrt{N_L}/\sqrt{(1+11.61x)}$~[ps]. To calculate the last expression
we used  the $a_0$ of GaAs ($a_0 = 0.565$~nm) also for the ternary
alloys, independent of $x$which is justified because of the weak
dependence of $a_0$ on In concentration: $a_0=0.565(1+0.072x)$~[nm]
[\onlinecite{Landolt}].

\begin{figure}[hbt]
\includegraphics[width=1.0\linewidth]{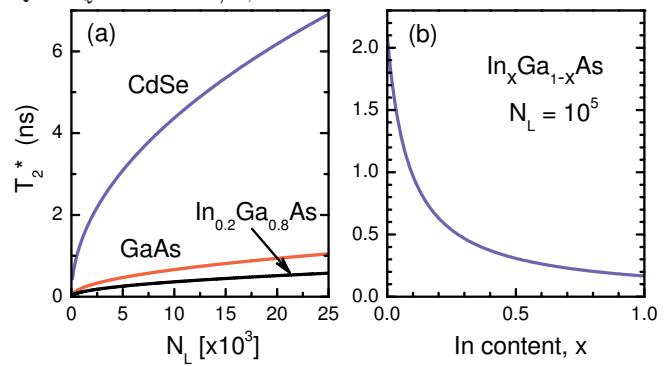}
\caption{(Color online) Model calculations of the nuclei fluctuation
effect on the electron spin dephasing time $T_2^*$:  (a) Dependence
of $T_2^*$ on the number of nuclei $N_L$ in the volume of electron
localization for CdSe, GaAs and In$_{0.2}$Ga$_{0.8}$As; (b)
Dependence of $T_2^*$ on In concentration in In$_x$Ga$_{1-x}$As for
$N_L=10^5$.} \label{fig:fig7}
\end{figure}

Figure~\ref{fig:fig7}(a) shows the calculated dependencies of the
electron spin dephasing time $T_2^*$ on the number of nuclei $N_L$
in the quantum dot, for CdSe, GaAs and In$_{0.2}$Ga$_{0.8}$As.
Figure~\ref{fig:fig7}(b) gives the dependence of $T_2^*$ on In
concentration in In$_x$Ga$_{1-x}$As for $N_L=10^5$.

The estimation of the number of nuclei $N_L$ in CdSe QDs (embedded
in a ZnSe matrix grown on a (001)-oriented GaAs substrate) is done
as follows. We assume cylinder-shaped pancake QDs so that volume is
given by QD $V_{QD}= \pi d^2h/4$, where $d$ and $h$ are diameter and
height of the QD, respectively. We assume that the QD has almost
zincblende crystal structure. Each cubic cell contains 4 Cd-Se
molecules, i.e. 8 atoms. Then the volume of the cell is $v_0=a_0^3$.
$N_L \approx 8V_{QD}/v_0$. For $d \sim 4-6$~nm and $h \sim
1.4-2.1$~nm one obtains $N_L \approx 630-2090$.

 \begin{widetext}
\begin{table} [h]
\caption{Parameters of nuclear isotopes used for calculating
electron spin dephasing times. Only isotopes with nonzero spin are
given here. All data for nuclear spins, $I_j$, and magnetic momenta,  $\mu_j$, are taken from Ref.~[\onlinecite{Handbook}]} \label{Table}
\begin{tabular}{|p{1cm}|p{1.5cm}|p{1.5cm}|p{2cm}|p{1.5cm}|p{1.5cm}|}
\hline
Species & $I_j$ & $\mu_j$ & Abundance $\varkappa_j$ & $\eta_j$, $\times 10^3$ & $\tilde{A_j}$, $\mu$eV \\
\hline
$^{111}$Cd & 1/2 & -0.5943 & 0.128 & 3.6 \footnote[1]{From Ref.~[\onlinecite{Nakamura79}]} & -37.4 \\
\hline
$^{113}$Cd & 1/2 & -0.6217 & 0.123 & 3.6 \footnotemark[1] & -39.1 \\
\hline
\hline
$^{77}$Se & 1/2 & +0.534 & 0.0758 & 3.6 & 33.6\\
\hline
\hline
$^{69}$Ga & 3/2 & +2.016 & 0.604 & 2.61 \footnote[2]{From Ref.~[\onlinecite{Paget77}]} & 38.2 \\
\hline
$^{71}$Ga & 3/2 & +2.562 & 0.396 & 2.61 \footnotemark[2] & 48.5 \\
\hline
\hline
$^{75}$As & 3/2 & +1.439 & 1 & 4.42 \footnotemark[2] & 46\\
\hline
\hline
$^{113}$In & 9/2 & +5.523 & 0.0428 & 6.35 \footnote[3]{From Ref.~[\onlinecite{Gueron}]} & 84.6 \\
\hline
$^{115}$In & 9/2 & +5.534 & 0.9572 & 6.35 \footnotemark[3] & 84.8 \\
\hline
\end{tabular}
\end{table}
\end{widetext}


\begin{thebibliography}{9}
\bibitem{comment1} The used form $\sim \exp(-t/T_2^*)\cos (\omega_L t)$ to fit the time evolution of the KR amplitude assumes that the resident electrons have a Lorentzian distribution of Larmor precession frequencies, e.g., due to a spread of $g$ factor. For a Gaussian distribution the following function should be used: $\sim \exp(-0.5(t /T_2^*)^2)\cos(\omega_L t)$. However, fits of the experimental data with both functions yield approximately the same estimates for $T_2^*$.
\end{thebibliography}
\end{document}